# Nanodosimetric investigation of the track structure of therapeutic carbon ion radiation. Part 1: Measurement of ionization cluster size distributions.

Gerhard Hilgers, Miriam Schwarze*, Hans Rabus

Physikalisch-Technische Bundesanstalt, Braunschweig and Berlin, Germany

* corresponding author, miriam.schwarze@ptb.de

**Abstract**

At the Heidelberg Ion-Beam Therapy Center, the track structure of carbon ions of therapeutic energy after penetrating layers of simulated tissue was investigated for the first time. Measurements were conducted with carbon ion beams of different energies and polymethyl methacrylate (PMMA) absorbers of different thicknesses to realize different depths in the phantom along the pristine Bragg peak. Ionization cluster size (ICS) distributions resulting from the mixed radiation field behind the PMMA absorbers were measured using an ion-counting nanodosimeter. Two different measurements were carried out: (i) variation of the PMMA absorber thickness with constant carbon ion beam energy and (ii) combined variation of PMMA absorber thickness and carbon ion beam energy such that the kinetic energy of the carbon ions in the target volume is constant. The data analysis revealed unexpectedly high mean ICS values compared to stopping power calculations and the data measured at lower energies in earlier work. This suggests that in the measurements the carbon ion kinetic energies behind the PMMA absorber may have deviated considerably from the expected values obtained by the calculations. In addition, the results indicate the presence of a marked contribution of nuclear fragments to the measured ICS distributions, especially if the carbon ion does not cross the target volume.

1. Introduction

Nanodosimetry focuses on investigating the physical characteristics of the microscopic structure of ionizing particle tracks. Track structure encompasses the sequence of interaction types and loci of a primary particle and all its secondaries, which reflects the stochastic nature of radiation interaction. The microscopic structure of the ionizing particle track is considered closely related to the biological effects of ionizing radiation. This is crucial for understanding the biological effects of ion beams [1–9], where the major fraction of radiation damage is mainly concentrated along and close to the primary ion trajectory or the trajectories of secondary recoil or fragment ions. In ion beam therapy, the knowledge of the microscopic track structure is relevant for determining relative biological effectiveness (RBE) in the spread-out Bragg peak (SOBP). Moreover, this knowledge is vital for predicting unwanted late effects of the treatment in irradiated healthy tissue, such as secondary cancer induction due to the exposition of the healthy tissue in the entrance channel of the ion beam.

For treatment planning of the dose administered to tumors in ion beam therapy, the local effect model [6,7,10,11] or the microdosimetric kinetic model [12–15] are generally used.

Other approaches have been proposed using either microdosimetric approaches for nanometric targets [16,17], DNA damage models based on nanodosimetry [18,19,5,20,21,9,22–31], or even more advanced models that include the effects of radiation-induced radical species [32,33]. Microdosimetry or nanodosimetry approaches have the advantage that the physical characteristics of the radiation field entering the models can be measured (in principle) or, at least, the simulation codes used for calculating the quantities of interest can be benchmarked by testing them on corresponding experiments.

While several groups [34–38] have reported measurements of microdosimetric quantities along pristine Bragg peaks or SOBPs of clinical carbon ion beams, this work reports the first such measurements using nanodosimetry. (Previous nanodosimetric investigations of carbon ion tracks concerned mono-energetic carbon ion beams with much lower energies at non-clinical accelerators [39–43]. The track imaging experiments of Laczko et al. [44] were for carbon ions with a mass per energy of 30 MeV/u. They achieved a resolution of 50 nm, that is, much larger than the few nm site sizes generally considered in nanodosimetry.) The measurements were performed in several beam time shifts at the Heidelberg ion-beam therapy center (HIT). The first set of experiments pertains to measurements at different depths in a phantom with a fixed energy of the incident carbon ion beam such that the results represent the variation of track structure characteristics along a pristine Bragg peak. In the second set of experiments, different combinations of carbon ion beam energy and depth in the phantom were employed and expected to produce the same energy of the carbon ions in the nanodosimeter, albeit with a different background of secondary heavy charged particles.

## 2. Materials and methods

*2.1 Nanodosimetric quantities*

In experimental nanodosimetry, the basic measuring quantity is the frequency distribution of the ionization cluster size (ICS), which represents a characteristic aspect of the ionization component of the track structure. The ICS is defined as the number $\nu$ of ionizations generated in a target volume by a primary particle and its secondaries. For reasons of simplicity, a cylindrical target volume is often regarded. A primary particle of radiation quality $Q$ (where $Q$ is determined by the particle type and its energy) can either traverse the target volume or pass it at a distance $d$ (impact parameter) from the longitudinal axis of the cylinder, as shown in Figure 1 for the case of an ion. The ICS produced in the target results from the superposition of the ionization component of the particle track structure and the geometric characteristics of the target volume. The ICS distribution is the statistical distribution of the probabilities $P_\nu(Q,d)$ of exactly $\nu$ ions being created in the target volume by the track of a primary particle of radiation quality Q passing at an impact parameter of $d$, normalized to unity according to Eq. (1).

$$\sum_{\nu=0}^{\infty} P_\nu(Q,d) = 1 \qquad (1)$$

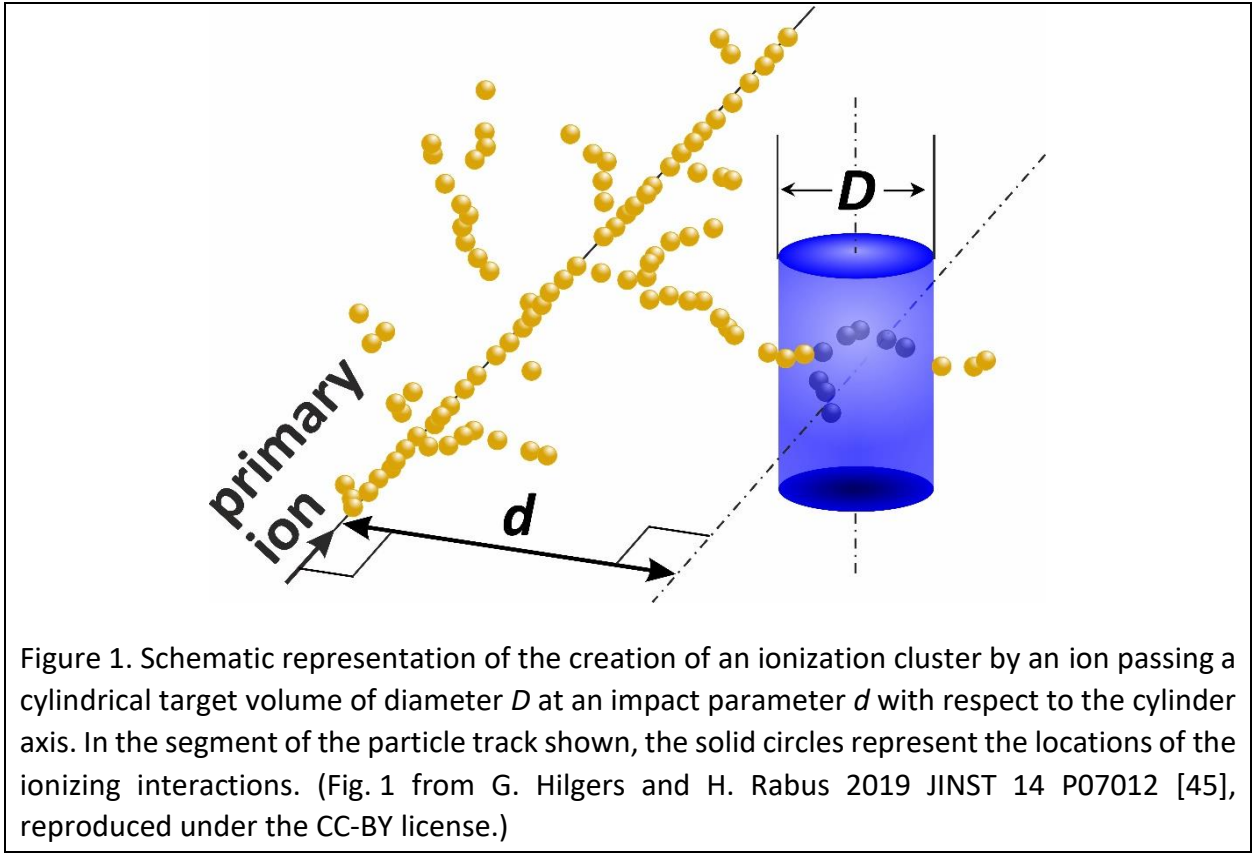

Figure 1. Schematic representation of the creation of an ionization cluster by an ion passing a cylindrical target volume of diameter *D* at an impact parameter *d* with respect to the cylinder axis. In the segment of the particle track shown, the solid circles represent the locations of the ionizing interactions. (Fig. 1 from G. Hilgers and H. Rabus 2019 JINST 14 P07012 [45], reproduced under the CC-BY license.)

Often, the mean ICS $M_1(Q,d)$ is of particular interest, which is defined by

$$M_1(Q,d) = \sum_{\nu=0}^{\infty} \nu \cdot P_\nu(Q,d) \ . \qquad (2)$$

The ICS distribution $P_\nu(Q,d)$ depends on the radiation quality $Q$ and the geometry of the target volume, its material composition and density, and the impact parameter $d$ of the charged particle trajectory concerning the target.

When the impact parameter is so large that the primary particle does not cross the target volume, the ICS distribution is dominated by events producing no ionizations, and the conditional ICS distribution may be more relevant. The conditional ICS distribution considers only events in which the primary particle track has generated at least one ionization in the target volume. The corresponding probabilities $P_\nu^C(Q,d)$ are given by Eq. (3).

$$P_\nu^C(Q,d) = \frac{P_\nu(Q,d)}{\sum_{\nu=1}^{\infty} P_\nu(Q,d)} \quad \text{for} \quad \nu \geq 1 \quad \text{with} \quad \sum_{\nu=1}^{\infty} P_\nu^C(Q,d) = 1 \qquad (3)$$

Consequently, the statistical moments $M_i^C(Q,d)$ of the conditional ICS distribution are defined in analogy to Eq. (2), such that $M_i^C(Q,d)$ is given by:

$$M_i^C(Q,d) = \sum_{\nu=1}^{\infty} \nu^i \cdot P_\nu^C(Q,d) \ . \qquad (4)$$

In previous investigations of monoenergetic carbon and helium ions of lower energies [39–41] found that the first three statistical moments $M_i^C(Q,d)$, i = 1 – 3, of the conditional ICS distribution were approximately constant (within statistical uncertainties on the order of 10 %) for large values of $d$, when the ion passes outside the target volume, and independent of radiation quality $Q$ (i.e., particle type and energy). This invariance of $M_i^C(Q,d)$ with impact parameter $d$ and its independence on the radiation quality $Q$ found were interpreted as showing that, at large impact parameter, the secondary electron spectrum changes only slightly with the radiation quality $Q$ and the distance of the target volume from the ion trajectory.

*2.2 Setup of the experiment*

The original setup of the experiment is extensively detailed in [46]. Later improvements regarding the data acquisition system, the data evaluation procedure, and improved characterization of the device are described in [47,45,48]. Figure 2 shows schematically the setup used in the present experiments and the idealized case that a carbon ion enters the nanodosimeter which is used in the following to illustrate the operation principle of the nanodosimeter.

The nanodosimeter comprises an interaction region filled with a rarefied target gas, an electrode system to extract target gas ions from the interaction region, an evacuated acceleration stage with an ion-counting detector at its end, and a position-sensitive particle detector. The interaction region is located between the electrodes of a plane parallel-plate capacitor and is filled with the target gas at a pressure of 1.2 hPa. An ionizing particle traversing the interaction region between the two electrodes produces target gas ions along its trajectory. The secondary particles produced in these interactions can produce further target gas ions in the vicinity of the trajectory.

The ionized target gas molecules generated by the traversing ionizing particle and its secondaries have thermal energies and drift toward the lower electrode due to the electric field applied across the plane parallel plate capacitor. This electric field is on the order of 10 V/cm and the mean free path between collisions is in the order of 10 µm so that the target gas ions are only transported without producing further ions on their way. Target gas ions passing through an aperture in the bottom electrode are extracted from the interaction region. Subsequently, the extracted target gas ions are transported through ion optics to an ion-counting secondary electron multiplier (SEM), where they are individually detected, and their arrival times recorded. The part of the vacuum system containing the ion optics is equipped with a differential pumping system to create a vacuum inside the section containing the SEM, ensuring a residual gas pressure suitable for operating the SEM.

Whether a target gas ion is extracted from the interaction region depends on the extraction efficiency at the position of the ion's creation. The extraction efficiency is rotationally symmetric around the central axis of the extraction aperture. It decreases with increasing radial distance from this axis and with increasing distance from the extraction aperture. The determination of the extraction efficiency through simulations and imaging is discussed in detail in [47,41]. $P_\nu(Q,d)$ is calculated from the measured data as $P_\nu(Q,d) = N_\nu/N_{tot}$, where $N_\nu$ is the number of events producing an ionization cluster of size $\nu$ and $N_{tot}$ is the total number of recorded events. The process of obtaining the cluster size from the measurement of a single event is described in detail in [46,47].

Data acquisition is started when an ionizing particle is registered by the position-sensitive particle detector (PSD) located behind the interaction region. The present experiment used two silicon strip detectors as one-dimensional PSDs, one in front and the other behind the interaction region, as shown in Figure 2. This is to enable the reconstruction of the primary particle´s trajectory. The active areas of the two PSDs are 2 mm in height and 10 mm in width (Sitek, 1L10, [49]). The PSDs are not pixel-based, but rather covered with resistive layers on the front side of the silicon chip contacted at the ends of the "length" axis. Position detection in the resistive layers works based on the charge division principle (for details see [41]). Thus, virtual pixels of arbitrary width can be configured in the off-line data processing. The uncertainties associated with the measurement of the ionization cluster size distribution (ICSD) and the target volume imaging due to the imaging properties of the PSD have been discussed in [41]. The centers of both PSDs were laterally shifted by 3 mm relative to the central axis of the target volume to allow a range for the impact parameter $d$ of up to 7 mm. Only events producing simultaneous signals in both PSDs were included in the off-line data analysis.

In addition to triggering the data acquisition, also pulse height spectra were recorded with the trigger detector (PSD behind the target volume). These pulse height spectra (Supplementary Figures S4 and S5) show a pronounced single peak, which corresponds to the carbon ions hitting the detector, and a continuum at channel numbers below the peak, which corresponds to secondary heavy charged particles belonging to the tracks of carbon ions, which miss the trigger detector or do not reach the nanodosimeter. To prevent triggering on these secondary particles, a window discriminator was applied in the off-line data processing, which covers only the carbon ion peak of the pulse height spectra. Discriminating only events with a carbon ion hitting the trigger detector does not lead to a loss of information on these events. This is because all target gas ions inside the target volume are collected irrespective of whether they are produced by the triggering carbon ion or any secondary particles of all generations belonging to the track of this carbon ion. On the contrary, the background due to secondary heavy charged particles belonging to tracks of carbon ions missing the trigger detector is suppressed for which an unambiguous determination of the impact parameter appears not possible.

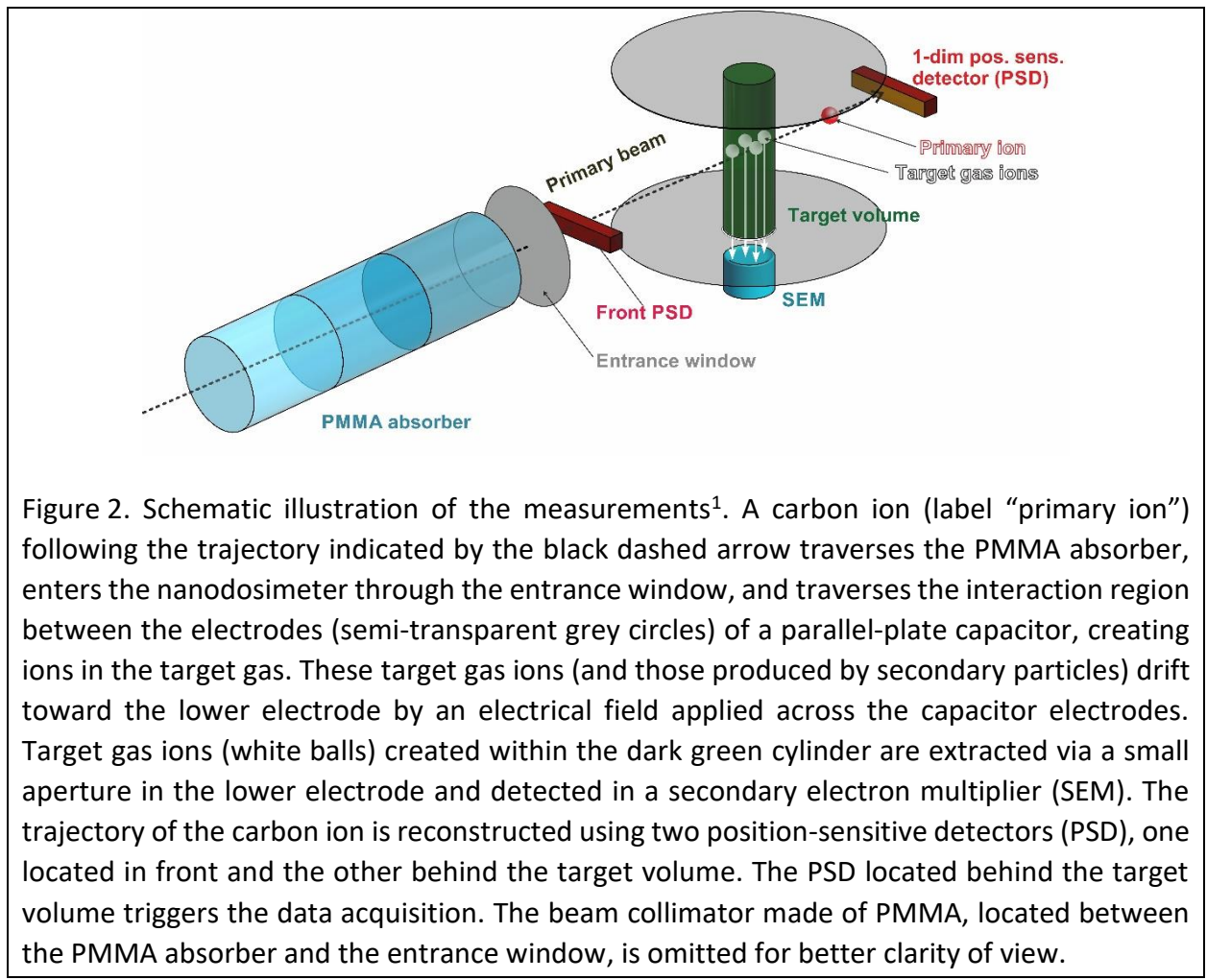

Figure 2. Schematic illustration of the measurements[1]. A carbon ion (label "primary ion") following the trajectory indicated by the black dashed arrow traverses the PMMA absorber, enters the nanodosimeter through the entrance window, and traverses the interaction region between the electrodes (semi-transparent grey circles) of a parallel-plate capacitor, creating ions in the target gas. These target gas ions (and those produced by secondary particles) drift toward the lower electrode by an electrical field applied across the capacitor electrodes. Target gas ions (white balls) created within the dark green cylinder are extracted via a small aperture in the lower electrode and detected in a secondary electron multiplier (SEM). The trajectory of the carbon ion is reconstructed using two position-sensitive detectors (PSD), one located in front and the other behind the target volume. The PSD located behind the target volume triggers the data acquisition. The beam collimator made of PMMA, located between the PMMA absorber and the entrance window, is omitted for better clarity of view.

The polymethyl methacrylate (PMMA) absorber was positioned about 50 cm from the beam exit of the beam line. About 45 cm downstream from the absorber, a collimator made of PMMA (not shown in Figure 2) with a thickness of 10 cm was placed about 1 cm from the entrance window of the nanodosimeter (5 mm A150 plastic). The collimator had an aperture with a height of 2 mm and a width of 10 mm; the bodies of the PMMA absorber and PMMA collimator had a square cross-section with sides measuring 30 cm each.

Carbon ions leaving the accelerator beam line and entering the PMMA absorber lose a portion of their kinetic energy during their passage through the PMMA absorber. To prevent confusion, the energy of carbon ions, which have left the beam line and not yet entered the PMMA absorber is referred to as "carbon ion beam energy" throughout the following text. For carbon ions in the nanodosimeter (having traversed the PMMA absorber), the energy is referred to as "carbon ion kinetic energy". From the perspective of the nanodosimeter, the carbon ions leaving the absorber are the primary particles of the nanodosimetric measurement.

---

[1] Note that the drawing is not to scale. The geometry of the nanodosimeter is described in detail in [46]. For a rough orientation: The capacitor plates have a diameter of 250 mm and are 50 mm apart. The dark green cylinder has a diameter of about 5 mm.

The measurements were performed using 1.2 hPa $C_3H_8$ as the target gas. A drift time window was applied such that only the target gas ions that reached the SEM between 73 μs and 129 μs after the registration of the carbon ion in the second PSD were counted. With this gas pressure and time window, the target volume corresponds to a cylinder with a diameter of about 1.0 mm and height of 5.5 mm, estimated with the methodology described [48]. According to the scaling procedure described in [5], this corresponds to a target cylinder in liquid water of between 2.9 nm and 3.0 nm in diameter (depending on the carbon ion energy). Using the same scaling factor of about 3 nm/mm, the maximum impact parameter of 7 mm would correspond to about 20 nm in liquid water. However, it must be noted that the scaling between different materials is based on the ratio of the quantities ($\rho\lambda_{ion}$) of the two materials, where $\rho$ is the mass density and $\lambda_{ion}$ is the mean free path for ionization by the ion in the respective material at this density. The product of the two quantities is equal to the ratio of the molecular mass of the target molecules and their cross section for ionization by the ion. Therefore, the scaling is dependent on the energy of the ion and, for the present experiments, strictly applies only to the carbon ion and not to its fragments or secondaries. For this reason, in this work all dimensions are given as they were in the experiment and are not scaled to liquid water at unit density.

Between the different beam time shifts, the whole setup was dismounted and completely removed from the beam line. Therefore, the first measurements in each beam time shift were performed under identical conditions. This served as a constancy check and allowed an estimate of the reproducibility and uncertainty of the measurements. The diameter of the carbon ion beam was 10 mm (full width at half maximum, corresponding to focus level 4), the repetition time of the beam pulse was 9 s with an extraction time of 5 s, a focus level of 4, and the dynamic intensity control (DIC) was switched off. The typical count rate of events was below 1000 $s^{-1}$ with a maximum count rate of up to 2000 $s^{-1}$.

Two different types of measurement were carried out: (i) variation of the PMMA absorber thickness with a constant carbon ion beam energy of 3.5 GeV (corresponding to an energy per mass of 292 MeV/u for $^{12}$C). (ii) combined variation of PMMA absorber thickness and carbon ion beam energy such that the kinetic energy of $^{12}$C ions inside the target volume of the nanodosimeter has the same value. During the preparation of the experimental setup, the energy loss of the carbon ion beam resulting from its passage through the PMMA absorber, entrance window and front PSD was determined through calculations using the SRIM code [50,51], assuming the beam to be isotopically pure $^{12}$C ions. The obtained kinetic energy values in the nanodosimeter target volume corresponding to the different absorber thicknesses in the first set of experiments are listed in Table 1. The corresponding estimated depths in water are also listed in Table 1. These were calculated by multiplying the PMMA thickness with a constant factor of 1.15837 ± 0.00042, determined from the ratios of the range in water to the range in PMMA of protons and alpha particles with energy per mass between 100 MeV/u and 400 MeV/u (Supplementary Figure S2). The corresponding data were retrieved from the PSTAR and ASTAR databases of the National Institute of Standards and Technology (NIST) [52].

Table 1. PMMA absorber thickness used in the experiments with a carbon ion beam energy of 3.5 GeV (energy per mass 292 MeV/u for $^{12}$C), the corresponding depth in water and the resulting kinetic energy of the carbon ions in the interaction volume of the nanodosimeter calculated with SRIM.

| PMMA thickness / mm | depth in water / mm | energy in target / GeV | energy per mass / MeV/u |
|---|---|---|---|
| 93  | 107.7 $\pm$ 0.1 | 1.82 $\pm$ 0.01 | 152 $\pm$ 1 |
| 111 | 128.6 $\pm$ 0.1 | 1.40 $\pm$ 0.02 | 117 $\pm$ 1 |
| 124 | 143.6 $\pm$ 0.1 | 1.02 $\pm$ 0.02 | 85 $\pm$ 2  |
| 129 | 149.4 $\pm$ 0.1 | 0.85 $\pm$ 0.02 | 70 $\pm$ 2  |
| 132 | 152.9 $\pm$ 0.1 | 0.72 $\pm$ 0.02 | 60 $\pm$ 2  |

Table 2. Carbon ion beam energies and PMMA absorber thicknesses used in the experiments. According to SRIM calculations, these combinations result in a kinetic energy of the carbon ions in the interaction volume of the nanodosimeter of 1 GeV (energy per mass ≈ 83 MeV/u for $^{12}$C).

| beam energy / GeV | energy per mass / MeV/u | PMMA thickness / mm | depth in water / mm |
|---|---|---|---|
| 2.5 | 208 | 60  | 69.5 $\pm$ 0.1  |
| 3.0 | 250 | 91  | 105.4 $\pm$ 0.1 |
| 3.5 | 292 | 124 | 143.6 $\pm$ 0.1 |
| 4.0 | 333 | 161 | 186.5 $\pm$ 0.1 |

## 3. Results

*3.1 Reproducibility and uncertainty of the experiments*

Since the whole setup (nanodosimeter, PMMA absorber, and collimator) was dismounted and completely removed from the beam line between the different beam time shifts, the first measurements in each beam time shift were performed with identical conditions (carbon ion beam energy, absorber thickness, target gas and pressure), thus serving as a constancy check and allowing to estimate reproducibility and uncertainty of the measurements.

Figure 3 shows $M_1(d)$ for $d \leq 7$ mm of the first measurement of each beam time shift. The measurements were performed in 1.2 hPa $C_3H_8$ with a carbon ion beam energy of 3.5 GeV ($\approx$ 292 MeV/u for $^{12}$C) and a PMMA absorber of 124 mm thickness. The $M_1(d)$ show good agreement except for the measurement in shift 2, where the $M_1(d)$ data for large $d$ deviate significantly from the other measurements. This deviation is due to an increased counting of background events in the SEM originating from an ionization vacuum gauge, which was not switched off at the beginning of the measurement. Consequently, the measurement in shift 2 is not included in determining the uncertainty. The relative uncertainty for $M_1(0)$ from these data is 5 % for a coverage factor $k$ = 2. For other values of $d$, the relative uncertainty of $M_1(d)$ ranges between 5 % and 10 %.

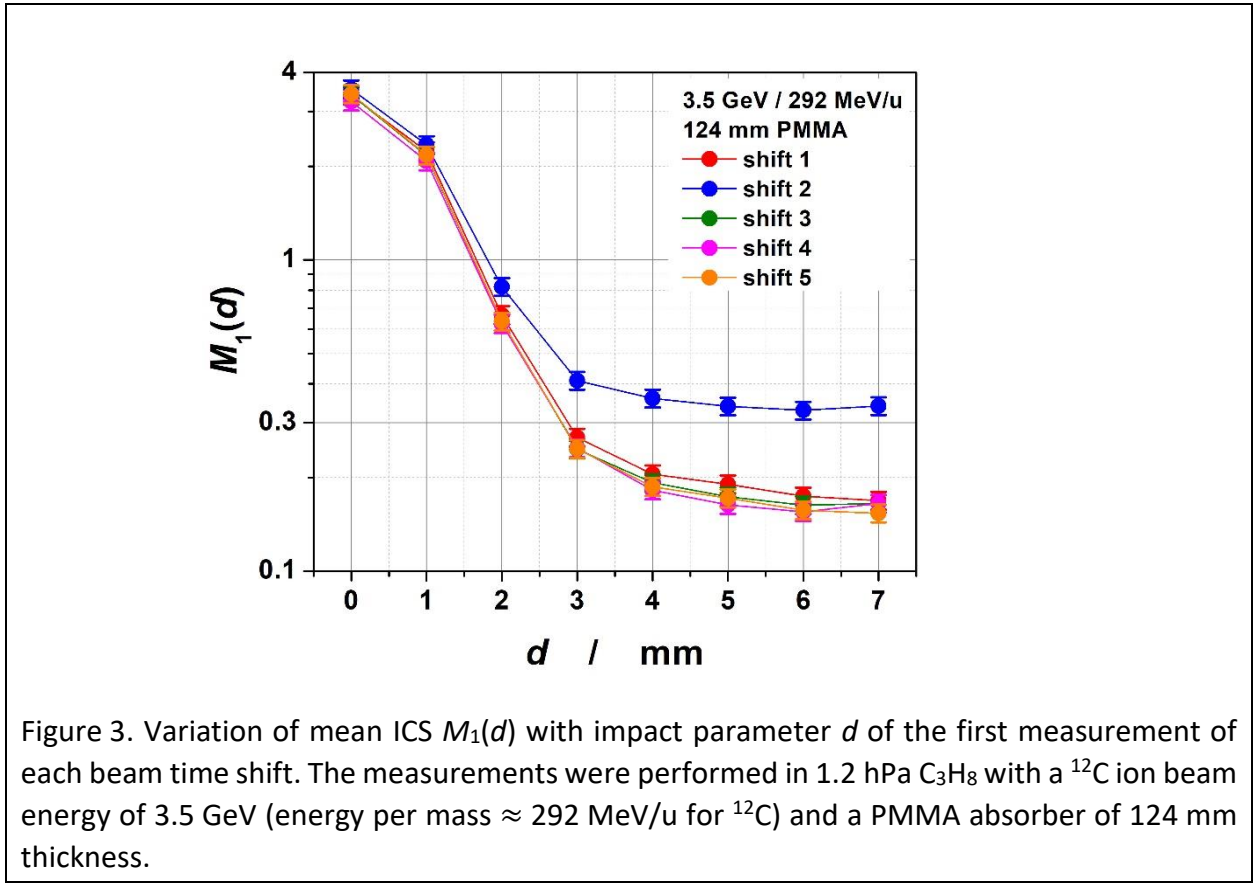

Figure 3. Variation of mean ICS $M_1(d)$ with impact parameter $d$ of the first measurement of each beam time shift. The measurements were performed in 1.2 hPa $C_3H_8$ with a $^{12}C$ ion beam energy of 3.5 GeV (energy per mass ≈ 292 MeV/u for $^{12}C$) and a PMMA absorber of 124 mm thickness.

*3.2 Variation of mean ICS and ICS distribution with PMMA absorber thickness*

The left plot of Figure 4 shows the relative frequency distribution of the ICS $P_\nu(d)$ for $d$ = 0 mm for the variation of the PMMA absorber thickness with a constant carbon ion beam energy of 3.5 GeV (energy per mass ≈ 292 MeV/u for $^{12}C$). The legend shows the PMMA absorber thickness and the kinetic energy of the $^{12}C$ ions in the target volume calculated with SRIM. With decreasing energy of the interacting carbon ions, the linear energy transfer (LET) of the carbon ions increases, and with it, the number of ionizations in the target volume. Therefore, the maximum in the ICS frequency distribution shifts towards larger ICS and the probability of creating an ionization cluster of large cluster size $\nu$ increases with decreasing energy.

The right plot in Figure 4 shows the ICS frequency distribution $P_\nu(d)$ for $d \leq 7$ mm for a carbon ion beam energy of 3.5 GeV and a PMMA absorber of 132 mm thickness, resulting in a kinetic energy of 0.72 GeV of the carbon ions (energy per mass ≈ 60 MeV/u for $^{12}C$) in the target volume. With increasing $d$, the maximum in the distribution shifts towards smaller ICS down to $\nu$ = 0, and the probability of producing an ionization cluster of large cluster size $\nu$ decreases. The reason for this behavior is, that for central passage of the target volume, i.e., for $d$ = 0 mm, the part of the carbon ion trajectory inside the target volume is at a maximum and decreases with increasing impact parameter $d$, i.e., the distance between the carbon ion's trajectory and the central axis of the target volume, leading to a decreasing number of ionizations in the target volume. At larger impact parameter $d$ ($d$ > 2 mm in the present setup), when the carbon ion's trajectory is completely outside the target volume, the changes in the ICS distributions are less pronounced than for smaller $d$, since at large $d$, the ionizations are exclusively due to secondary particles. Further examples of ICS distributions are shown in Supplementary Figure S3.

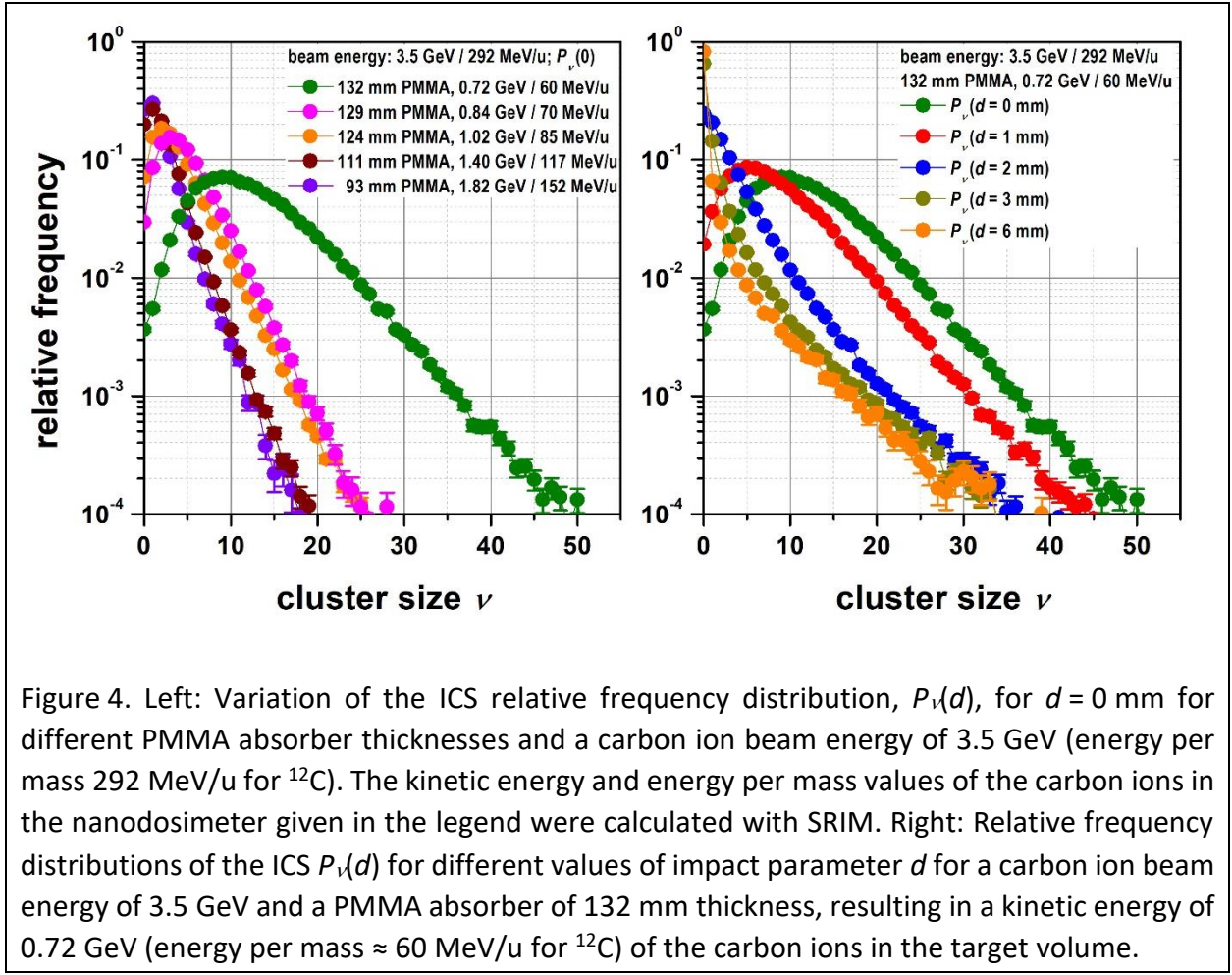

Figure 4. Left: Variation of the ICS relative frequency distribution, $P_\nu(d)$, for $d = 0$ mm for different PMMA absorber thicknesses and a carbon ion beam energy of 3.5 GeV (energy per mass 292 MeV/u for $^{12}$C). The kinetic energy and energy per mass values of the carbon ions in the nanodosimeter given in the legend were calculated with SRIM. Right: Relative frequency distributions of the ICS $P_\nu(d)$ for different values of impact parameter $d$ for a carbon ion beam energy of 3.5 GeV and a PMMA absorber of 132 mm thickness, resulting in a kinetic energy of 0.72 GeV (energy per mass ≈ 60 MeV/u for $^{12}$C) of the carbon ions in the target volume.

Figure 5 shows the mean ICS $M_1(d)$ for $d \leq 7$ mm calculated with Eq. (2) for the variation of PMMA absorber thickness with a constant carbon ion beam energy of 3.5 GeV (energy per mass ≈ 292 MeV/u for $^{12}$C). As in Figure 4, the legend shows the PMMA absorber thickness, the kinetic energy of the carbon ions in the target volume of the nanodosimeter (calculated with SRIM), and the corresponding energy per mass. The $M_1(d)$ data for the different absorber thicknesses appear to shift toward a larger mean ICS with increasing PMMA absorber thickness. This is expected as increasing the thickness of the PMMA absorber means increasing LET of the carbon ions leaving the absorber due to their decreasing remaining kinetic energy. This results in increasing $M_1(d)$ values with increasing PMMA absorber thickness.

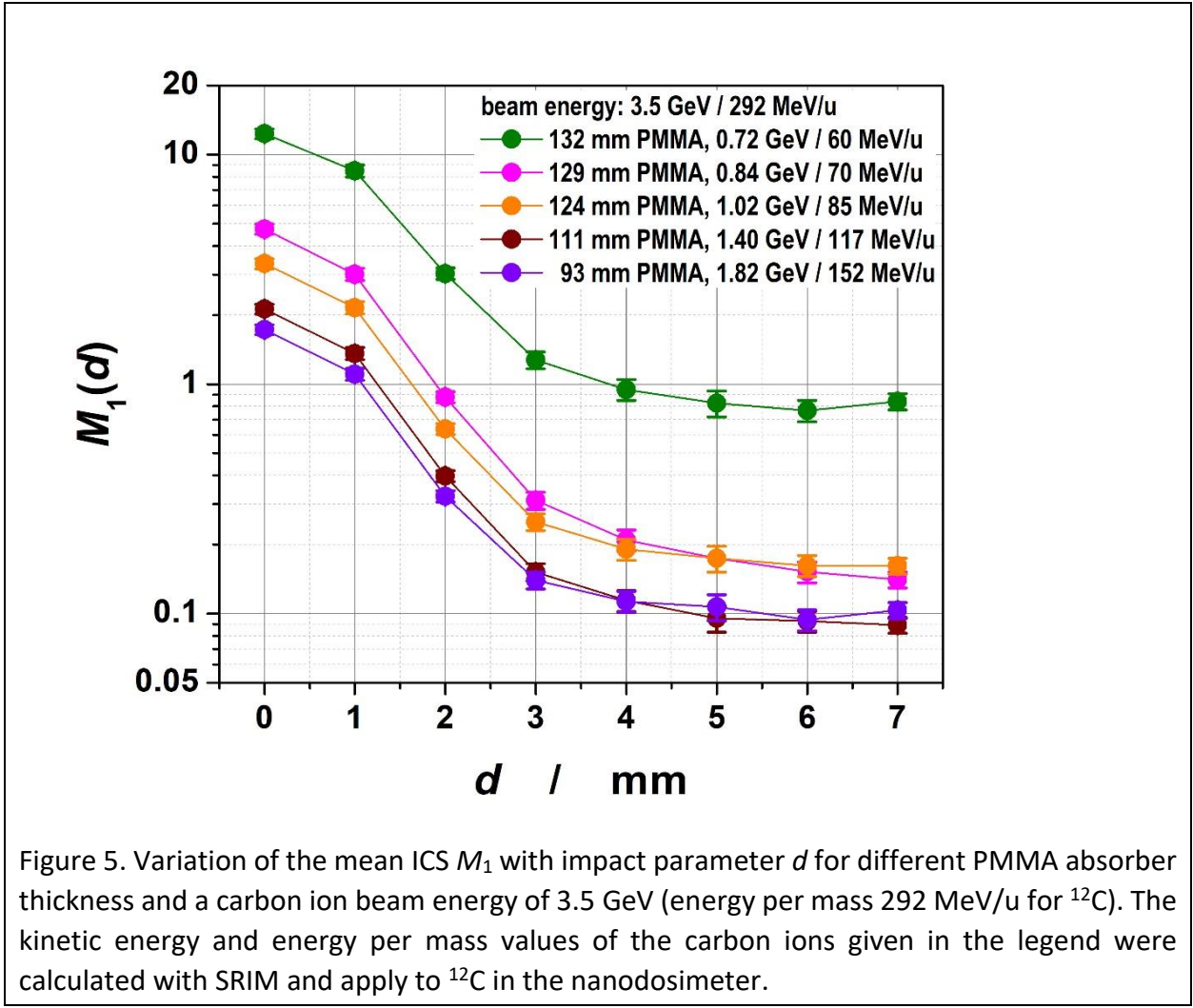

Figure 5. Variation of the mean ICS $M_1$ with impact parameter $d$ for different PMMA absorber thickness and a carbon ion beam energy of 3.5 GeV (energy per mass 292 MeV/u for $^{12}$C). The kinetic energy and energy per mass values of the carbon ions given in the legend were calculated with SRIM and apply to $^{12}$C in the nanodosimeter.

A constant shift along the logarithmic vertical axis of Figure 5 corresponds to a constant factor on a linear scale. However, the variation of the $M_1(d)$ curves is not only by a constant factor as illustrated in Supplementary Figure S1. Here, the data were normalized to the respective values at $d = 0$ mm. Particularly for the lowest residual kinetic energy of 0.72 GeV (energy per mass of 60 MeV/u for $^{12}$C), an increase in the ratio $M_1(d)/M_1(0)$ can be observed. Notably, a scaling of $M_1(d)$ with $M_1(0)$ is not expected since for increasing $d$, the relative contribution to $M_1(0)$ from ionizations due to secondary particles increases, and the composition of the secondary radiation field cannot be assumed to be independent of the PMMA absorber thickness [53].

Figure 6 shows the first (i.e., the conditional mean ICS) and second moment of $P_\nu^C(Q,d)$ obtained from the present measurements using Eq. (4). For both moments of the conditional ICS distribution, a pronounced variation in the values at the larger impact parameter with absorber thickness can be observed, which amounts to several tens of percent for $M_1^C(d)$ and up to a factor in the order of 10 for $M_2^C(d)$. In addition, $M_2^C(d)$ has a pronounced variation with increasing impact parameter, especially for the data of 132 mm absorber thickness (note that the vertical axis of $M_2^C(d)$ in Figure 6 is logarithmic). These findings are at variance with what was observed in earlier investigations of monoenergetic carbon and helium ions of lower energies [39–41], where both quantities were found to be constant within about 10 % as a function of the impact parameter and also for different radiation qualities.

At large distances (i.e., $d > 2$ mm in the present setup) from the carbon ion's trajectory, when the carbon ion passes outside the target volume, the ionization of target gas molecules in the target volume is exclusively due to secondary particles. In contrast to the mentioned earlier work, the carbon ions passed through a PMMA absorber in the present experiments. Therefore, ionizations in the target volume from a carbon ion track passing at large impact parameter are not only due to secondary electrons produced by the carbon ion but also by tracks of other heavy charged particles belonging to the same event.

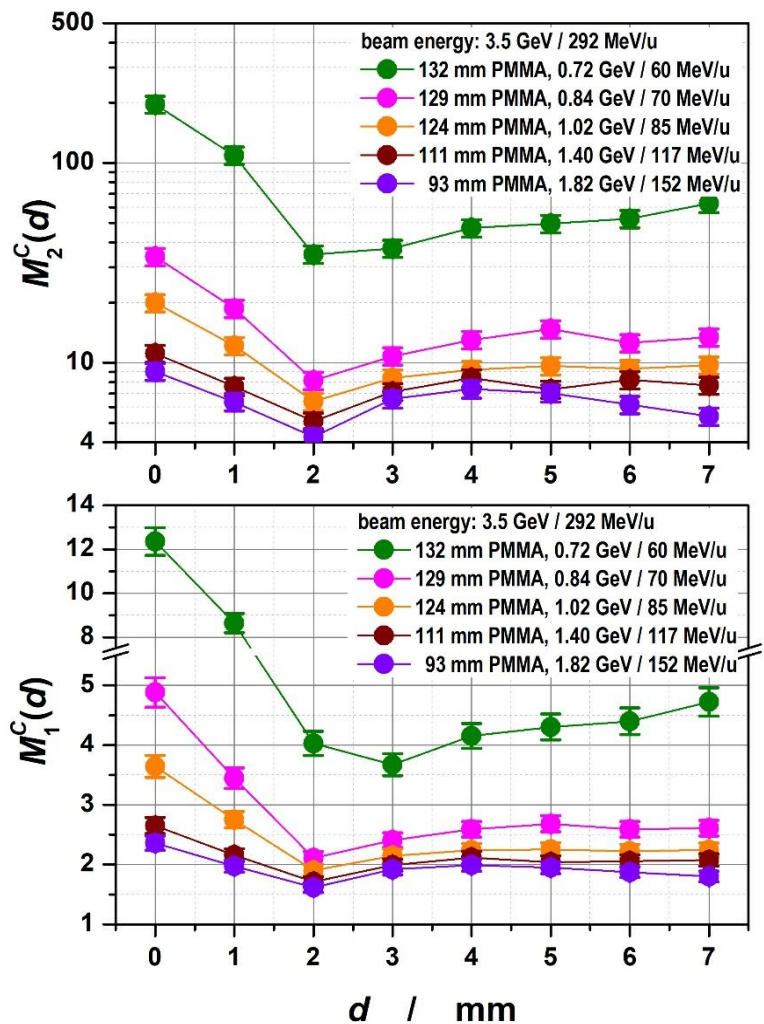

Figure 6. Conditional mean ICS $M_1^C(d)$ and second moment $M_2^C(d)$ of the conditional ICS distribution with the impact parameter $d$ for the variation of PMMA absorber thickness with constant carbon ion beam energy of 3.5 GeV (≈ 292 MeV/u energy per mass for $^{12}$C). The kinetic energy and energy per mass values of the carbon ions in the legend were calculated for $^{12}$C ions using SRIM.

*3.3 Comparison with measurements at lower energies*

The different dependence of track structure characteristics on impact parameter for measurements with and without a PMMA absorber is confirmed by Figure 7, which shows a comparison of the mean ICS $M_1(d)$ for $d \leq 7$ mm obtained in this study and measurements without a PMMA absorber, both having the same $M_1(0)$ of 12.3. The measurements without absorber were performed during the BioQuaRT (Biological Quantities in Radiation Therapy) project [41]. The data pertaining to the measurements with the absorber shows a significantly larger $M_1(d)$ for large $d$ than that without the absorber, suggesting an increasing number of ionizations due to secondaries. In addition, for this combination of 3.5 GeV carbon ion beam energy and absorber thickness of 132 mm, it appears that the conditional first two moments show a pronounced increase with increasing impact parameter (Figure 6) whereas $M_1(d)$ appears approximately constant for $d \geq 4$ mm in Figure 7.

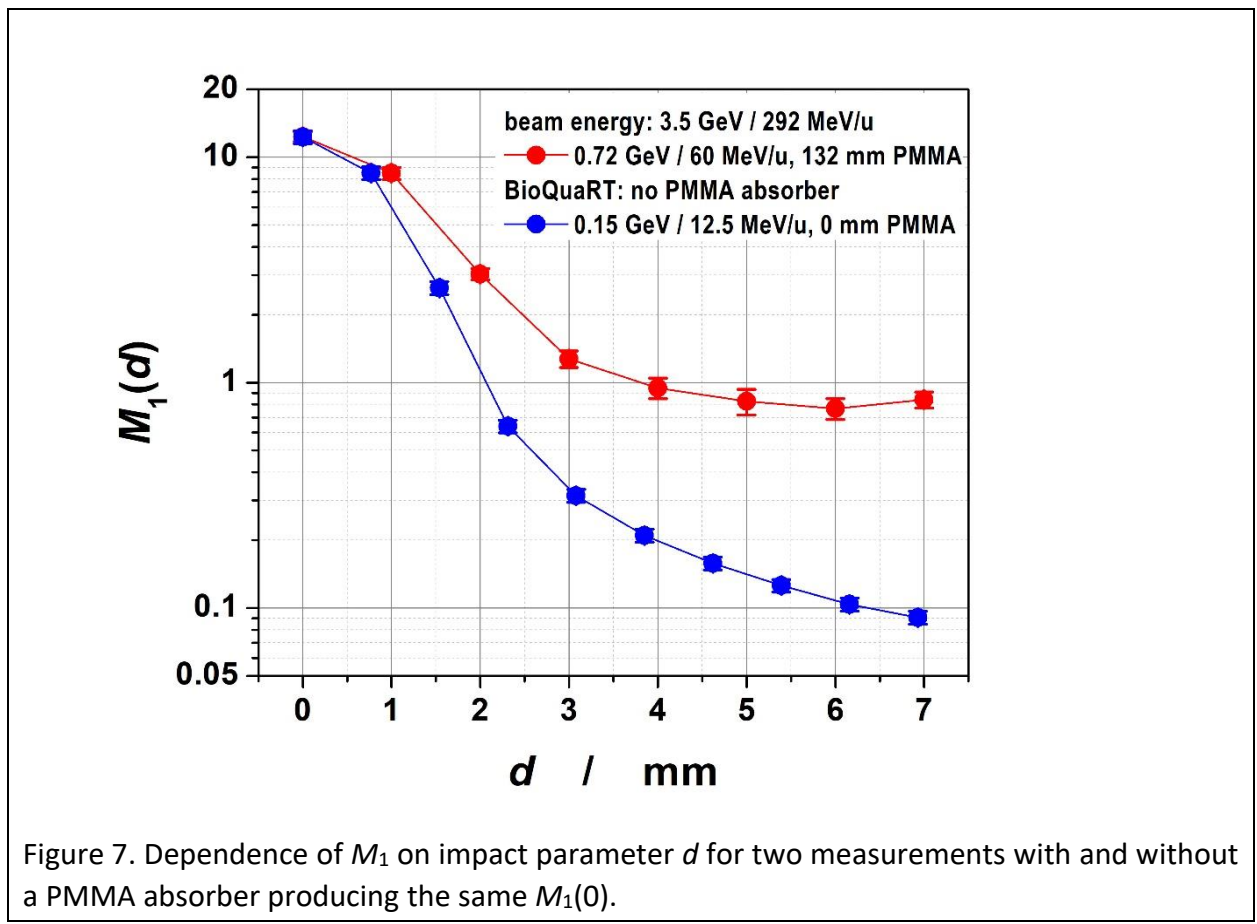

Figure 7. Dependence of $M_1$ on impact parameter $d$ for two measurements with and without a PMMA absorber producing the same $M_1(0)$.

Figure 8 compares the mean ICS for the central passage of the carbon ion through the target, $M_1(0)$, between the present data and the results of previous measurements without a PMMA absorber performed during the BioQuaRT project [41]. In addition, the red symbols (referring to the vertical axis on the right-hand side) show the mass stopping power $S(E)/\rho$ of $^{12}$C ions in $C_3H_8$ calculated with SRIM. Measurements without a PMMA absorber generally follow the relative energy dependence of the mass stopping power. For measurements with a PMMA absorber, increasing deviations from the curve of $S(E)/\rho$ are found with increasing thickness of the PMMA absorber toward larger $M_1(0)$, indicating an increasing number of ionizations due to secondaries with high LET.

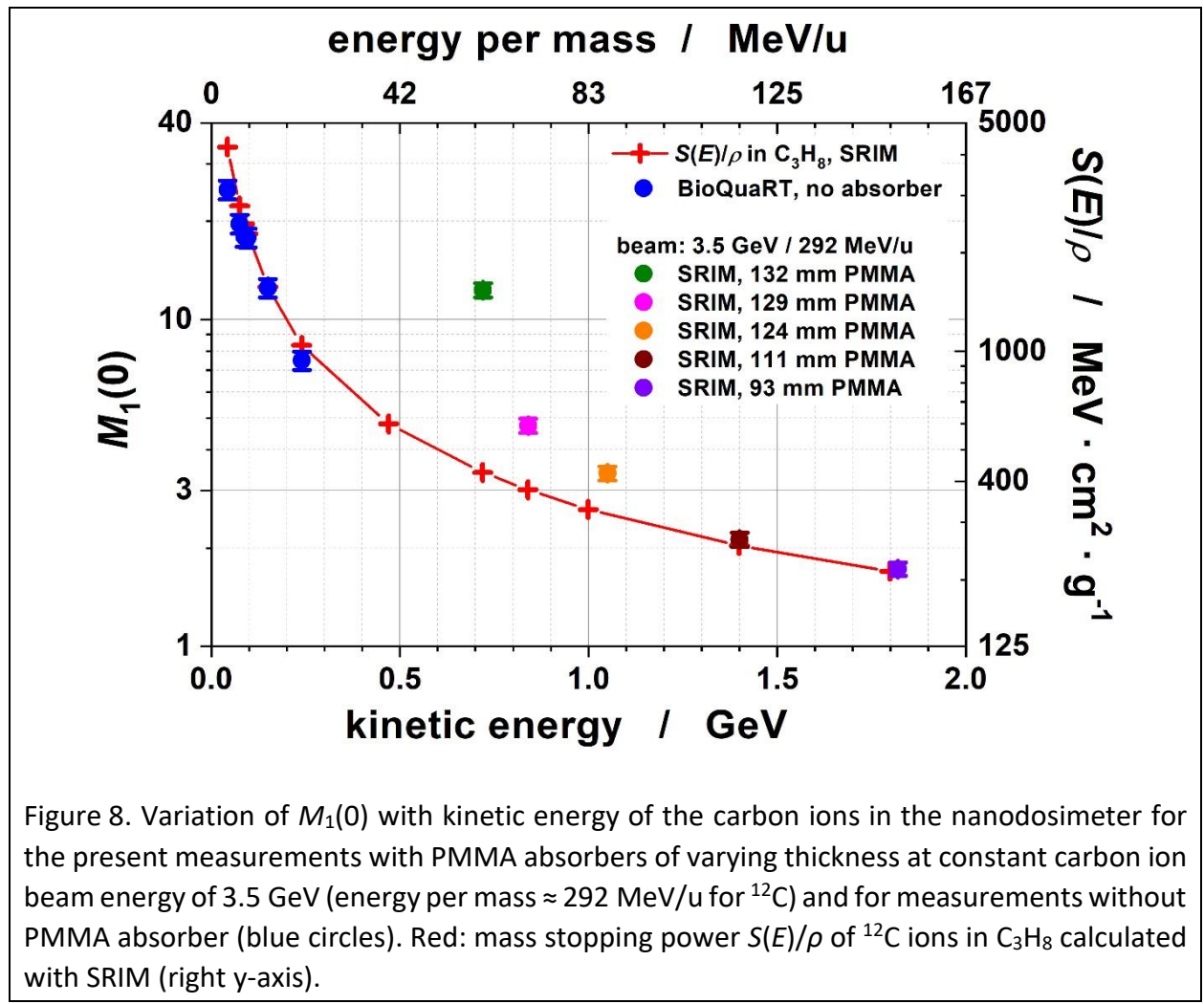

Figure 8. Variation of $M_1(0)$ with kinetic energy of the carbon ions in the nanodosimeter for the present measurements with PMMA absorbers of varying thickness at constant carbon ion beam energy of 3.5 GeV (energy per mass ≈ 292 MeV/u for $^{12}$C) and for measurements without PMMA absorber (blue circles). Red: mass stopping power $S(E)/\rho$ of $^{12}$C ions in $C_3H_8$ calculated with SRIM (right y-axis).

While the relation between the mass stopping power (pertaining to the total ionization produced per mass) and the track structure (spatial distribution of ionization clusters) is not obvious a priori, the heuristic comparison between the two quantities is motivated by the left plot of Figure 9, which shows a comparison between the $M_1(0)$ values and the peak channel number of the signal recorded with the PSD behind the target volume, which serves as trigger detector. The right plot of Figure 9 shows the ratio of $M_1(0)$ to the channel of the peak in the pulse height spectrum. (For the PMMA absorber thickness of 132 mm, the mean of the pulse height values corresponding to 50 % of the maximum is shown instead, see text below.) The peak channel number is proportional to the number of ionizations produced and the energy imparted by an impinging carbon ion in the PSD. (The carbon ions are not completely stopped in the PSDs due to their small thickness and they deposit only a small fraction of their kinetic energy in the detector.)

The peak channel number shifts towards larger channel numbers with increasing PMMA absorber thickness with the same relative dependence on PMMA absorber thickness as $M_1(0)$ (Figure 9 left). For all PMMA absorber thicknesses investigated, the $M_1(0)$ to peak channel number ratio is constant within the (large) uncertainties estimated from the full width at half maximum of the pulse height distributions shown in Supplementary Figure S4. As can be seen in Supplementary Figure S4, the width of the pulse height distribution increases with increasing PMMA absorber thickness to such an extent, that for an absorber thickness of 132 mm, the distribution becomes a broad plateau structure covering the upper half of the pulse height spectrum.

The kinetic energy of the carbon ions (energy per mass) in the target volume ranges between 0.72 GeV (60 MeV/u) and 1.82 GeV (152 MeV/u) (Table 1) for the given PMMA absorber thicknesses. According to SRIM calculations, the ratio of the mass stopping power of $^{12}$C ions in silicon to that of $^{12}$C ions in $C_3H_8$ varies only by about 3.5 % in the range between 0.5 GeV (42 MeV/u) and 2 GeV (167 MeV/u) kinetic energy (energy per mass for $^{12}$C). Therefore, Figure 9 suggests that the mean ionization cluster size $M_1(0)$ produced by carbon ions in $C_3H_8$ and the mass stopping power $S(E)/\rho$ of $^{12}$C ions in $C_3H_8$ have a similar energy dependence. In addition, Supplementary Figure S4 indicates that the energy distribution of the carbon ions in the nanodosimeter increases significantly in width with increasing PMMA absorber thickness.

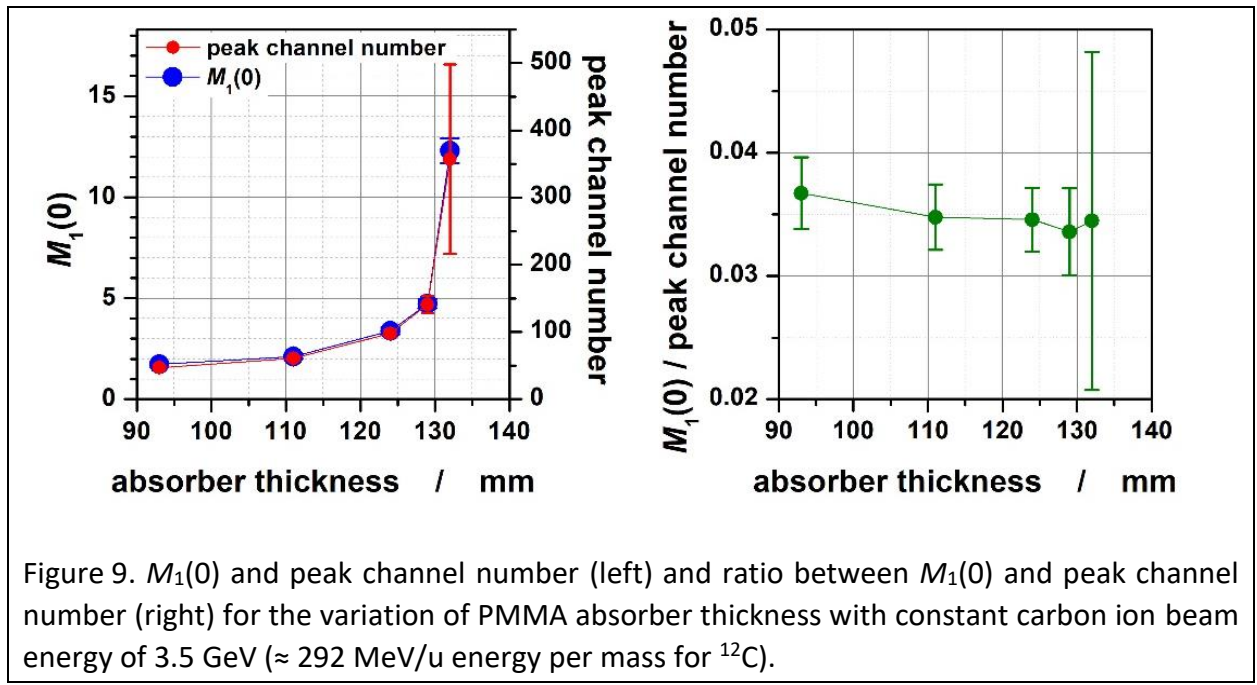

Figure 9. $M_1(0)$ and peak channel number (left) and ratio between $M_1(0)$ and peak channel number (right) for the variation of PMMA absorber thickness with constant carbon ion beam energy of 3.5 GeV (≈ 292 MeV/u energy per mass for $^{12}$C).

*3.4 Combined variation of PMMA absorber thickness and primary carbon ion energy*

The left plot of Figure 10 shows the mean ICS $M_1(0)$ for $d = 0$ mm for the combined variation of PMMA absorber thickness and carbon ion beam energy such that the kinetic energy of the carbon ions interacting inside the target volume is constant. The targeted kinetic energy in the target volume was 1 GeV (energy per mass ≈ 83 MeV/u for $^{12}$C). The thickness required for the PMMA absorber was obtained from calculations with SRIM of the energy loss of $^{12}$C ions due to the PMMA absorber, entrance window and front PSD. $M_1(0)$ is found to increase with increasing carbon ion beam energy and increasing thickness of the PMMA absorber. This is confirmed by the corresponding $M_1(d)$ for $d ≤ 7$ mm shown in the right plot of Figure 10, which are shifted towards larger mean ICS with increasing beam energy. A possible reason for this behavior could be an increase of the number of secondary particles of high LET with increasing carbon ion beam energy and increasing thickness of the PMMA absorber.

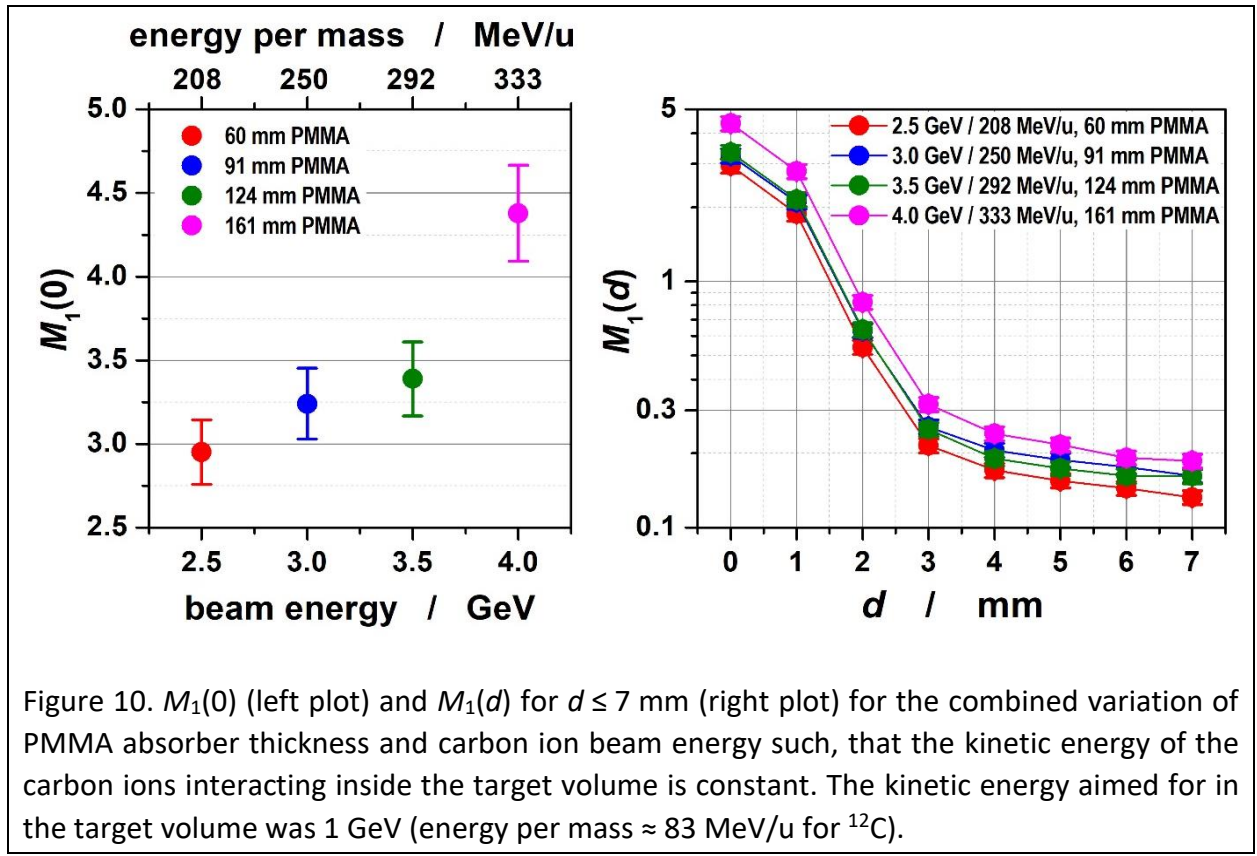

Figure 10. $M_1(0)$ (left plot) and $M_1(d)$ for $d \leq 7$ mm (right plot) for the combined variation of PMMA absorber thickness and carbon ion beam energy such, that the kinetic energy of the carbon ions interacting inside the target volume is constant. The kinetic energy aimed for in the target volume was 1 GeV (energy per mass ≈ 83 MeV/u for $^{12}$C).

Another reason might be an incorrect calculation of the energy loss by SRIM, as reported in earlier work [54,55]. This is supported by the pulse height spectra recorded with the PSD behind the target volume serving as the trigger detector, shown in Supplementary Figure S5. In the off-line data analysis, the peak corresponding to the energy deposition of the carbon ions was found to shift towards larger channels with increasing kinetic energy and PMMA absorber thickness (left plot in Figure 11), indicating an increasing energy loss in the detector and thus an increasing LET, which leads to an increasing mean ionization cluster size. The right plot of Figure 11 shows the ratio of $M_1(0)$ to the channel of the peak in the pulse height spectrum. For the combinations of PMMA absorber thickness and the carbon ion beam energy investigated, the ratio of $M_1(0)$ to peak channel is constant within the uncertainties, indicating a strict correlation between the two quantities. This, in turn, supports the assumption of an incorrect calculation of the energy loss in the PMMA absorber by SRIM, which might also affect the energies assigned to the measurements with therapeutic carbon ions shown in Figure 8.

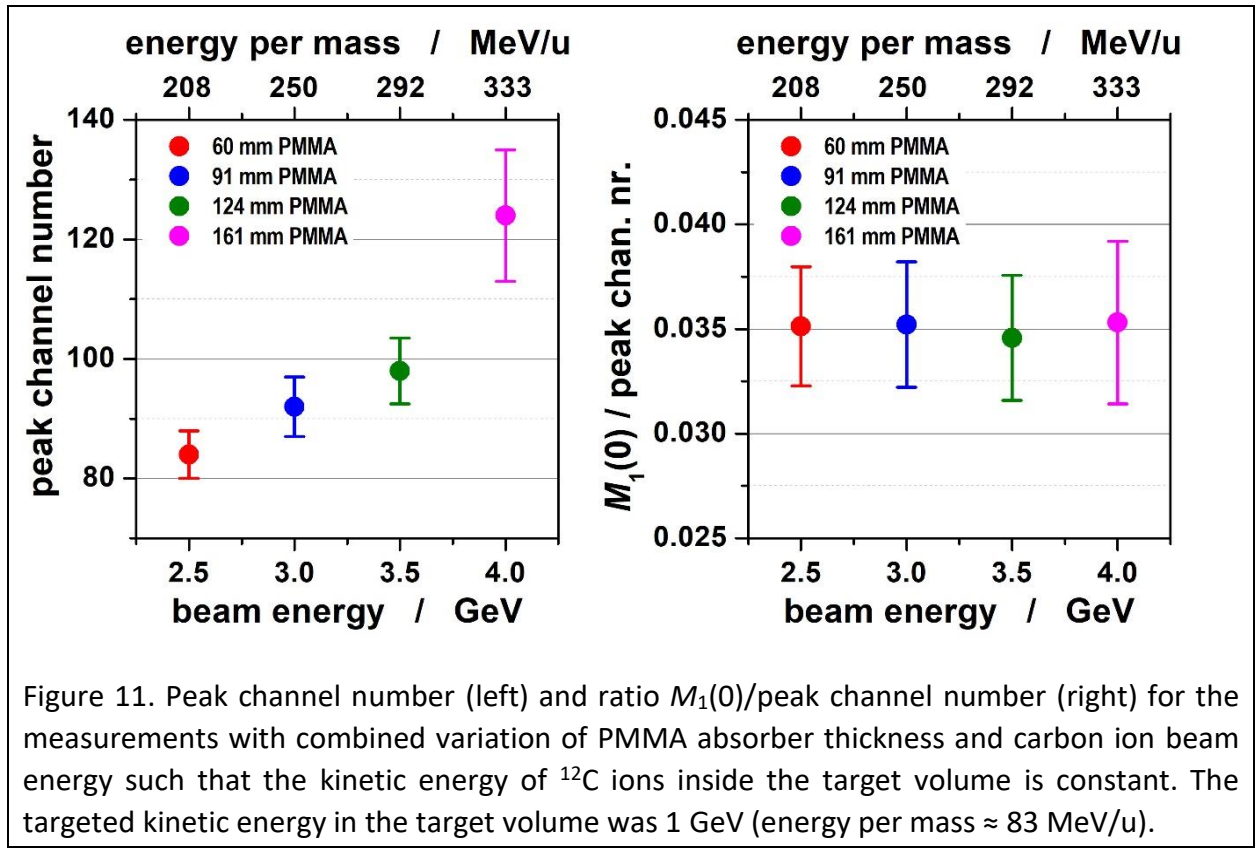

Figure 11. Peak channel number (left) and ratio $M_1(0)$/peak channel number (right) for the measurements with combined variation of PMMA absorber thickness and carbon ion beam energy such that the kinetic energy of $^{12}$C ions inside the target volume is constant. The targeted kinetic energy in the target volume was 1 GeV (energy per mass ≈ 83 MeV/u).

## 4. Discussion and Conclusions

Nanodosimetric concepts in particle therapy treatment planning have been explored in previous studies [16,27,56–60] and are currently under intensive investigation. Recently it gained new interest with the development of a more profound theoretical basis [30,31]. The mentioned approaches rely on numerical methods (track structure simulations), which require the support of suitable experiments for benchmarking. This study contains the first nanodosimetric measurements at a clinical carbon ion beam behind absorbers representing human tissue and producing a mixed radiation field. The ultimate goal of the nanodosimetric approach is to characterize the radiation quality of such mixed fields [24].

In this context, the present experiments are pioneering and concentrated on the radiation field near the axis of the carbon ion beam. For a more comprehensive characterization of the radiation field, further measurements at angles up to several degrees from this axis should be performed in future work to also characterize the nanodosimetric effects of the secondary heavy charged particles, which have a corresponding spread of their angular distribution [53]. Despite being focused on the vicinity of the carbon ion beam and the measurements with the nanodosimeter being reproducible within a few percent between the different beam time shifts at HIT, the present work yielded surprising results. The variation of the ICS distribution along a pristine Bragg peak of a 3.5 GeV (292 MeV/u for $^{12}$C) carbon ion beam showed several unexpected outcomes. First, the mean ICS at the noncentral passage of the carbon ions relative to the target appears enhanced compared to the measurements of "clean" (monoenergetic) carbon ion beams (Figure 7). Contrary to what was previously observed for monoenergetic beams, the first moments of the conditional ICS distribution were found to vary with the impact parameter and the radiation quality in the present experiments (Figure 6). In addition, the energy dependence of the mean ICS produced for the central passage of the target by the carbon ions showed a significant increase beyond what would be expected based on earlier work and calculated mass stopping powers using the SRIM code (Figure 8). Finally, measurements with combinations of the carbon ion beam energy and PMMA absorber thickness that should give the same carbon ion energy in the nanodosimeter were found to give significantly different absolute magnitude of the mean ICS (Figure 10).

For these latter measurements, the peak in the pulse height spectra of the PSD shifts towards larger channels with increasing beam energy and PMMA absorber thickness (Figure 11). This peak height corresponds to the carbon ions' energy deposition in the detector and depends on their kinetic energy. Therefore, Figure 10 and Figure 11 suggest a potential problem when calculating the energy loss of the carbon ions in the PMMA absorber using SRIM because SRIM predicted the same mean kinetic energy of the carbon ions in the nanodosimeter for these combinations of absorber thickness and beam energy. The large deviations of the present results for the mean ICS for carbon ions passing the target centrally (Figure 8) corroborate this conjecture. The problem is presumably due to a lack of consideration of nuclear processes or inappropriate cross-sections used for these.

On the other hand, the ratio of $M_1(0)$ and peak channel number of the signal was found to be constant within the (large) uncertainties, suggesting an approximate proportionality between the two quantities (Figure 9, Figure 11). It must be noted, however, that such a proportionality is not expected as such, since $M_1(0)$ is a parameter related to track structure, whereas the energy loss in the PSD is related to the total number of ionizations. In addition, it should be emphasized that the approximate proportionality between the two parameters indicated by Figure 9 and Figure 11 does not inform the choice of the implicit scaling factor used in Figure 8, which was conveniently chosen such that the stopping power curve matched the data point at the highest energy. An alternative choice would have been to have the stopping power curve fit the data at low energies, in which case the data points representing the new data all deviated from the stopping power curve.

Turning to the discrepancies found when the carbon ions do not pass through the target volume, it is worth reminding that for monoenergetic carbon and other ion beams, the first three moments of the conditional ICS distribution were constant within about 10 % for different radiation qualities at impact parameters for which the ionization clusters were only due to secondary particles [39–41]. Here, variations between several tens of percent and up to an order of magnitude (between different radiation qualities) were observed (Figure 6).

One explanation for this observation (and the enhanced mean ICS for carbon ions passing outside the target volume, Figure 7) could be the presence of a significant background of ionizations from secondary heavy charged particle tracks. It is known from a number of investigations, for instance, the one by Haettner et al. [53], that carbon ion beams exhibit markable contamination by nuclear fragments, which increases with increasing depth in water. This does not imply that the secondary heavy charged particles (1st generation secondaries) pass through the target volume. Secondary electrons (2nd generation secondaries) of these heavy charged particles of sufficiently large ranges may produce ionizations contributing to the detected ICS even if the heavy charged particle passed outside the target volume.

The enhanced mean ICS at impact parameters for which the carbon ion does not cross the target volume could also be due to a potential bias of the impact parameter determined from the interpolation of the center of gravity of the ionizations produced in the PSDs and the actual impact parameter of a carbon ion passing the target volume. Such a bias could result if the secondary heavy charged particles from the carbon ion track impacted the trigger detector together with the carbon ion. To affect the determination of the impact parameter from the PSD measurements, the secondary heavy charged particles must arrive with a time offset pertaining to the carbon ion smaller than the rise time of the PSD signal, which is in the order of a few tens of nanoseconds.

Since the secondary heavy charged particles have energies per mass comparable to the carbon ions [53], they also have a speed in the order of half the speed of light and traverse the whole experimental setup within a time in the order of 3 ns. Therefore, it cannot be ruled out that secondary heavy charged particles from a carbon ion track contribute to the signals in the two PSDs. However, it must be considered that the second PSD subtends a solid angle seen from the end of the PMMA absorber of only 0.7° × 0.14° or $2.5 \times 10^{-6}$ sr.

Haettner et al. [53] reported helium ions as the most abundant secondary heavy charged particles in carbon ion beams travelling in water. They found a particle radiance of helium ions in a forward direction of below 40 $sr^{-1}$ per carbon ion for a 400 MeV/u carbon ion beam and 159 mm depth in water. This depth corresponds approximately to the PMMA absorber thickness of 132 mm, but the carbon ion beam energy in our experiments was 3.5 GeV (energy per mass 292 MeV/u for $^{12}C$). The work of Haettner et al. [53] suggests that the total yield of helium ions is about a factor of 5 smaller for 200 MeV/u carbon ions. Therefore, for the case of 3.5 GeV beam energy (energy per mass 292 MeV/u for $^{12}C$) and 132 mm PMMA thickness, one could expect a helium particle radiance in the order of magnitude 20 $sr^{-1}$ per carbon ion. This resulted in $5 \times 10^{-5}$ helium ions hitting the trigger detector per carbon ion.

However, not all carbon ions reach the second PSD. According to Haettner et al. [53], the number of $^{12}C$ ions decreases approximately exponentially with depth in water, whereby the linear attenuation coefficient is the same for different initial energies of the $^{12}C$ ions. At a depth in water of 152.9 mm about 55 % of the carbon ions remain. Assuming the carbon ion beam to initially have a negligible divergence, the fraction of carbon ions hitting the second PSD in our experiment can be estimated at about 7.8 %. Here, the 55 % attenuation and the 10 mm FWHM of the beam (assumed to be rotationally symmetric) were considered. Therefore, the coincident arrival of helium ions with a carbon ion at the second PSD should occur with a probability of about $6 \times 10^{-4}$ and thus be negligible. It is understood that the values estimated above can only be rough approximations. However, preliminary simulation results of the present experiment [61] indicate that they are not unrealistic and that, in particular, deviations between actual impact parameters and those determined from the energy deposits in the PSDs larger than 0.05 mm occur only at a frequency below 0.1 % (Supplementary Figure S5 in [61]).

Given the broad distributions of the pulse height spectra shown in Supplementary Figure S4, one might wonder whether the measurements could also contain events in which the measurement is triggered by secondary heavy charged particles other than carbon ions. In fact, the additional peaks at the left end of the spectra may be due to, for instance, helium ions. When their energy per mass is similar to that of the carbon ion, their expected LET would be an order of magnitude smaller owing to the proportionality with the square of the nuclear charge. This is compatible with the position of the low energy peaks relative to the main peaks in Supplementary Figure S4.

In this context it is worth mentioning that Supplementary Figure S4 also indicates that the energy distribution of the carbon ions hitting the second PSD is much broader than predicted by the calculations with SRIM (cf. Table 1). To better understand this observation and the other measured results, a Monte Carlo simulation of the experiment was developed after the data analysis of the experiments was completed. The results of these simulations will be reported in the second part of the paper.

**Acknowledgments**

The authors gratefully acknowledge the developers of the nanodosimeter from the Weizmann Institute of Science, Rehovot, Israel, for transferring the device as described in [46] to the German National Metrology Institute (PTB) for further use. The authors also express their gratitude to B. Lambertsen and A. Pausewang for their invaluable contributions to the preparation and carrying out of the measurements and their assistance in data processing. The authors would like to thank S. Brons from HIT for his assistance and support in the preparation of the measurements and the staff of the ion accelerator facilities at HIT for their assistance and support during the measurements.

5. **Supplement**

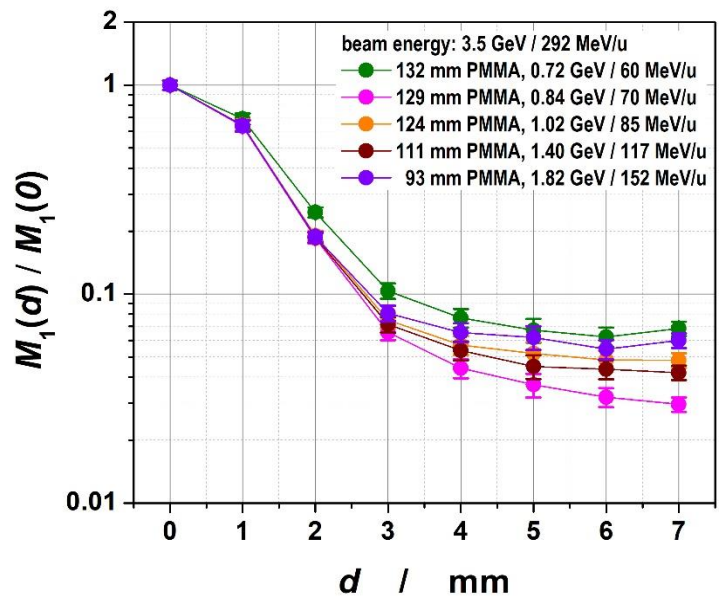

Supplementary Figure S1: Relative variation of the mean ICS $M_1$ with impact parameter $d$ for different PMMA absorber thickness and a carbon ion beam energy of 3.5 GeV (energy per mass 292 MeV/u for $^{12}$C).

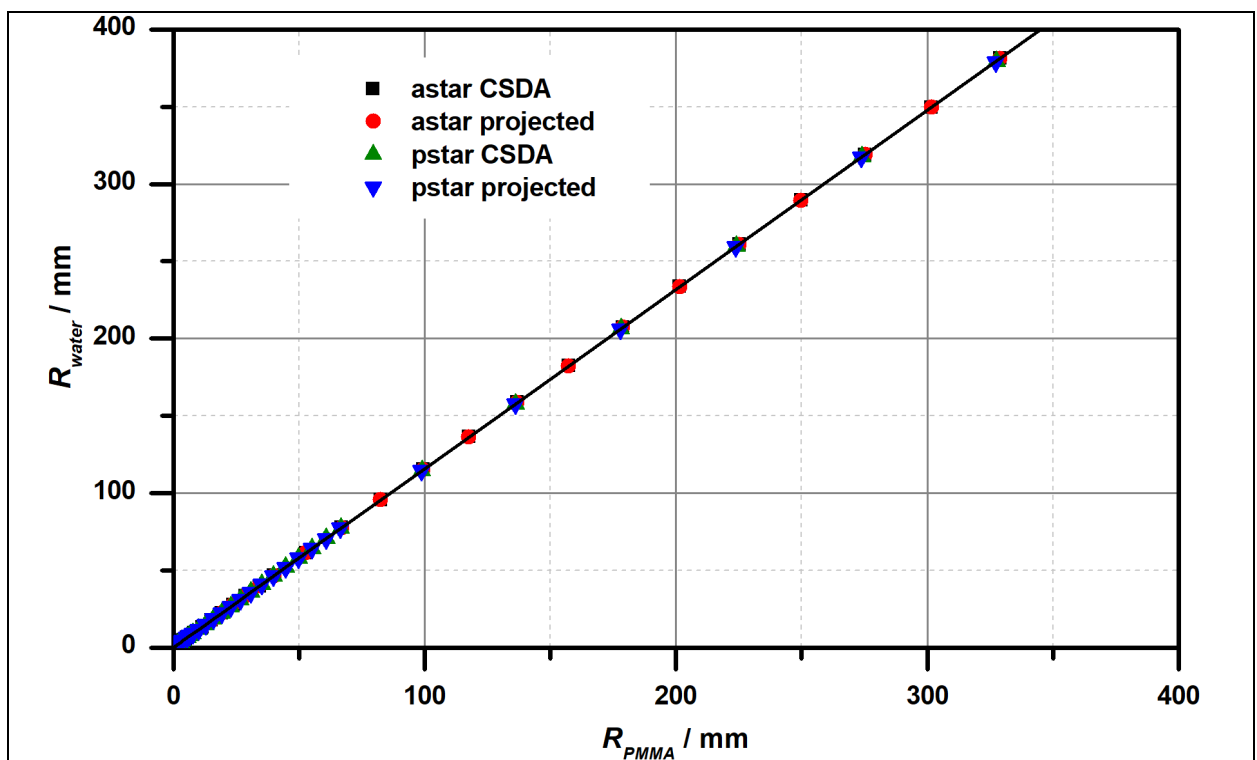

Supplementary Figure S2: Range in water versus range in PMMA of protons and alpha particles (values at the same respective energies taken from the NIST PSTAR and ASTAR databases [52]). The solid black line is a straight line passing through the origin presenting the best fit to the data corresponding to energies per mass between 100 MeV/u and 400 MeV/u. The linear regression gives a slope of 1.15837 ± 0.00042. This conversion factor is also used for carbon ions, assuming that ions with the same energy per mass have the same range.

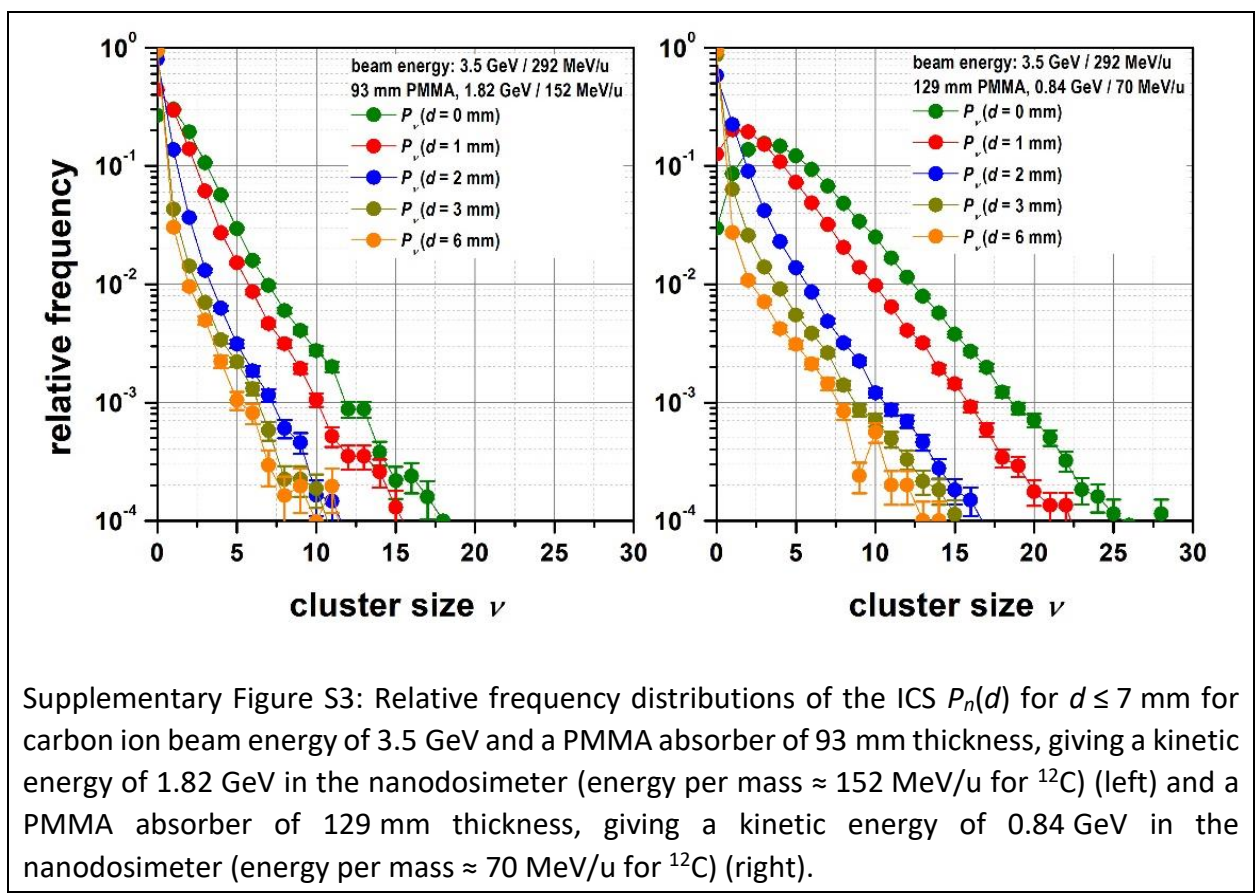

Supplementary Figure S3: Relative frequency distributions of the ICS $P_n(d)$ for $d \leq 7$ mm for carbon ion beam energy of 3.5 GeV and a PMMA absorber of 93 mm thickness, giving a kinetic energy of 1.82 GeV in the nanodosimeter (energy per mass ≈ 152 MeV/u for $^{12}$C) (left) and a PMMA absorber of 129 mm thickness, giving a kinetic energy of 0.84 GeV in the nanodosimeter (energy per mass ≈ 70 MeV/u for $^{12}$C) (right).

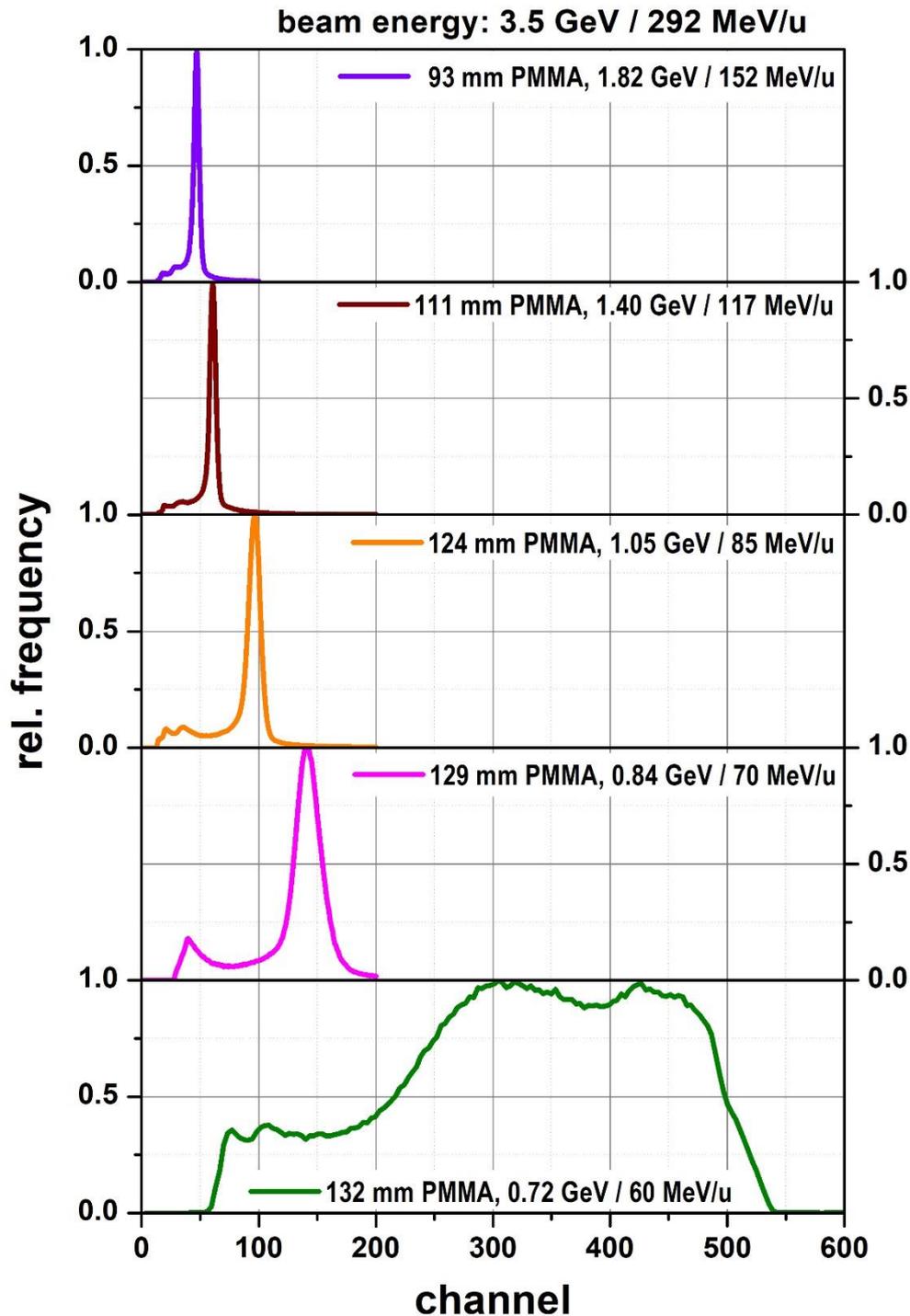

Supplementary Figure S4: Pulse height spectra recorded with the PSD behind the target volume for a carbon ion beam energy of 3.5 GeV (energy per mass ≈ 292 MeV/u for $^{12}$C) and PMMA absorbers of different thicknesses. For better comparability the spectra are normalized with respect to the maximum peak height. The channel numbers have been corrected for the gain of the preamplifier, which varied between the experiments with different absorber thicknesses. Since measurements are triggered when the preamplifier output exceeds the fixed discriminator threshold, the pulse height spectra start at different lowest channel numbers.

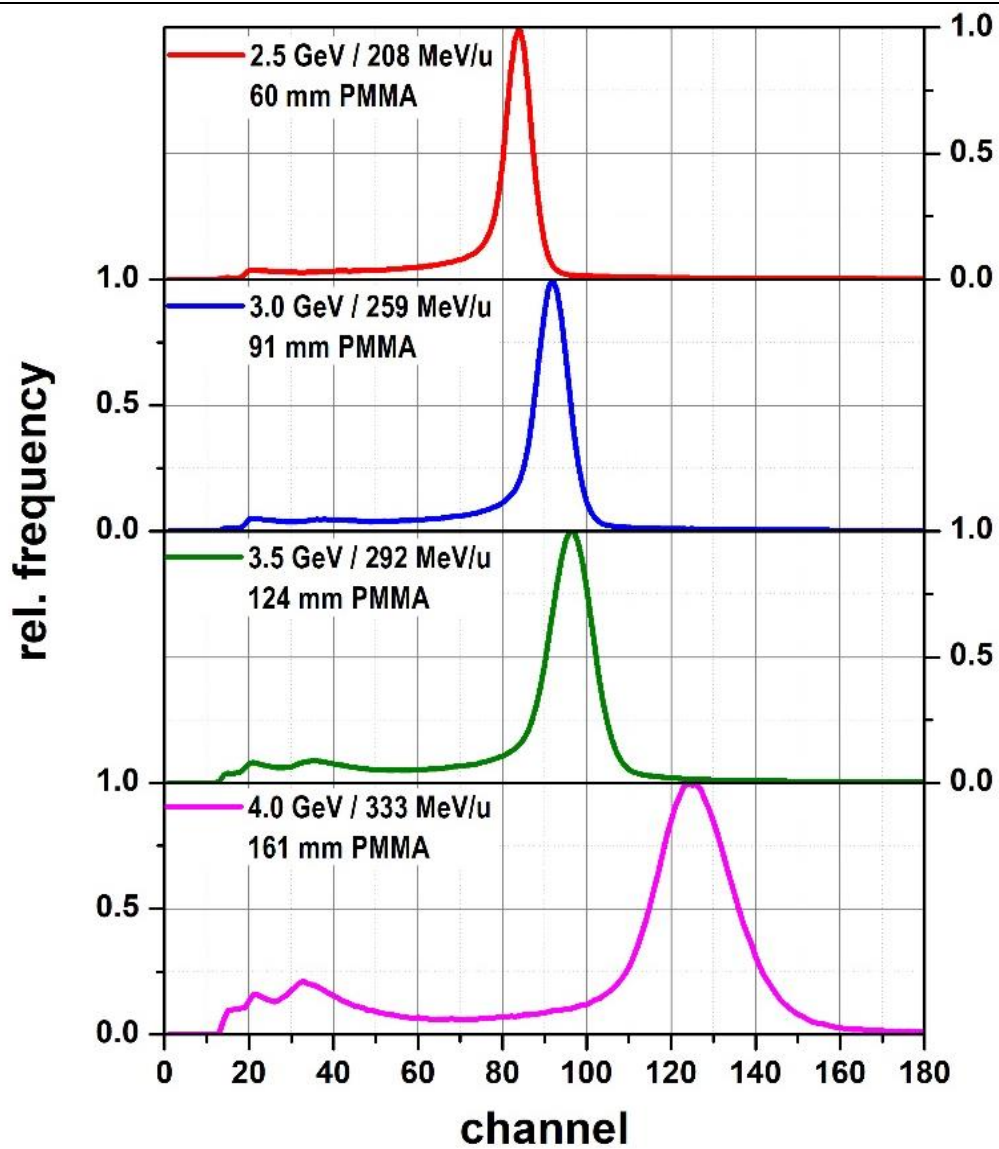

Supplementary Figure S5: Pulse height spectra recorded with the PSD behind the target volume for the combined variation of PMMA absorber thickness and carbon ion beam energy such that the mean energy of a carbon ion in the target volume is constant. The targeted total kinetic energy in the target volume was 1 GeV (energy per mass ≈ 83 MeV/u for $^{12}$C). For better comparability the spectra are normalized with respect to the maximum peak height.